# Magnetoresistive detection of perpendicular switching in a magnetic insulator

Silvia Damerio[1,*], Achintya Sunil[2], M. Mehraeen[2], Steven S.-L. Zhang[2] and Can O. Avci[1,*]

[1]Institut de Ciència de Materials de Barcelona, Campus de la UAB, Bellaterra, 08193, Spain

[2]Department of Physics, Case Western Reserve University, Cleveland, Ohio 44106, USA

**Spintronics offers promising routes for efficient memory, logic, and computing technologies. The central challenge in spintronics is electrically manipulating and detecting magnetic states in devices. The electrical control of magnetization via spin-orbit torques is effective in both conducting and insulating magnetic layers. However, the electrical readout of magnetization in the latter is inherently difficult, limiting its use in practical applications. Here, we demonstrate magnetoresistive detection of perpendicular magnetization reversal in an electrically insulating ferrimagnet, terbium iron garnet (TbIG). To do so, we use TbIG|Cu|TbCo, where TbCo is a conducting ferrimagnet and serves as the reference layer, and Cu is a nonmagnetic spacer. Current injection through Cu|TbCo allows us to detect the magnetization reversal of TbIG with a simple resistance readout during an external magnetic field sweep. By examining the effect of measurement temperature, TbCo composition, and Cu thickness on the sign and amplitude of the magnetoresistance, we conclude that the spin-dependent electron scattering at the TbIG|Cu interface is the underlying cause. Technologically-feasible magnetoresistive detection of perpendicular switching in a ferrimagnetic garnet is a breakthrough, as it opens broad avenues for novel insulating spintronic devices and concepts.**

Ferrimagnetic garnets (FMGs) are ubiquitous in solid-state physics. They host a wide variety of useful properties, such as ultralow damping, high magnon density, tunable magnetic anisotropy and saturation magnetization, and highly ordered single-crystal structures with sharp surfaces and interfaces.[1] Furthermore, single-crystal epitaxial FMGs can be grown from a few micrometers down to a few nanometers of thickness via physical and chemical deposition techniques on suitable substrates and, yet, retain bulk-like properties. As a result, FMGs have served as an excellent material testbed for a multitude of spintronic phenomena in the past few years.[2–8] More recently, there has been a growing effort in exploiting FMGs as active components in magnonic integrated circuits[9] and spintronic memory devices.[10] Research into the latter has been particularly fruitful owing to the development of ultrathin FMGs with perpendicular magnetic anisotropy (PMA)[11,12] and the highly efficient control of their magnetization vector by current-induced spin-orbit torques.[8,10,13–17]

While the above advances are promising for the integration of FMGs in future computing technologies, the electrical detection of magnetization in this material family is still a critical challenge. The spin Seebeck, local and non-local spin Hall magnetoresistance, and thermal spin drag effects have been utilized for the electrical detection of magnetization in FMGs.[3–5,18] However, all of these phenomena rely on voltages induced by inverse spin Hall effect in an adjacent nonmagnetic metal, such as Pt, hence fundamentally limiting the output signal amplitudes and materials that can be used for the purpose. Electrical detection of the perpendicular magnetization vector in FMGs possessing PMA, crucial for domain wall and skyrmion-based racetrack devices, and for downscaling the lateral size of memory cells, is particularly strenuous. Thus far, the anomalous Hall effect (AHE) driven by interfacial spin-dependent scattering[4,19–21] is used as the sole electrical method for magnetization vector detection in PMA FMGs,[10,22] yielding minute output signals, nonetheless in a technologically-prohibitive four-contact Hall cross device geometry.

Magnetization vector detection in conventional spintronic devices is realized by tunnel and giant magnetoresistances (GMR).[23–26] The latter, discovered in the 1980s, has led to miniaturized GMR sensors, revolutionizing both the magnetic non-volatile data storage and field sensing technologies.[27] What lies at the core of the GMR is the spin valve effect, arising from the spin-dependent electron scattering at the bulk and interfaces of ferromagnetic materials.[28] GMR spin valves typically consist of conducting magnetic layers separated by nonmagnetic metal spacers. The electrical resistance of such systems exhibits a large asymmetry between the parallel and antiparallel alignment of the magnetizations of adjacent magnetic layers, which serve as the reference and free layers, respectively. GMR can then be used to effectively probe the magnetization reversal or rotation of the free layer with respect to the reference, also in a current-in-plane (CIP) geometry, leading to simple two-terminal devices. To date, FMGs have been excluded from such device concepts even though the GMR-based sensing of FMGs' magnetization vector would be highly desirable. Few earlier reports have characterized the spin valve effect at other magnetic insulator|metal interfaces, but these were limited to in-plane magnetization systems and cryogenic temperature measurements.[29–31] While the presence of the spin valve effect is allowed at an FMG|metal interface in principle, its experimental observation remains elusive.

In this work, we show a robust spin valve effect at the interface of a typical FMG, terbium iron garnet ($Tb_3Fe_5O_{12}$, TbIG), enabling us to detect 180º perpendicular magnetization reversal via simple resistance measurements at room temperature. We achieve such spin valve behavior in a TbIG|Cu|TbCo trilayer (see Fig. 1a) where the TbIG is engineered to be the free layer whereas TbCo acts as the reference magnetic layer. We observe abrupt longitudinal resistance changes during a swept magnetic field upon crossing the coercive fields of the respective layers, analogous to the CIP-GMR, albeit with a smaller amplitude with respect to a fully metallic stack. The effect of measurement temperature, TbCo layer composition, and spacer layer thickness on the sign and amplitude of the magnetoresistance collectively pinpoint the spin valve effect at the TbIG|Cu interface as the predominant origin of the observed phenomenon. Theoretical calculations based on layer-resolved Boltzmann transport equations are in excellent agreement with our experimental findings and further consolidate the presumed origin of the magnetoresistance effect. The efficient reading of the magnetization vector in FMGs in a two-terminal CIP geometry reported here provides a new platform for novel spintronic device concepts using magnetic insulators as active components.

**Layer composition and magnetic characterization**

We deposited TbIG(25)|Cu(1-5)|Tb$_x$Co$_{1-x}$(8)|Pt(3) (thicknesses in nm, subscript $x$ denotes atomic %) stacks and other reference layers by means of magnetron sputtering, and patterned them into standard Hall bar devices (see Fig. 1b and Methods). Here, TbIG and Tb$_x$Co$_{1-x}$ (TbCo for short, $x = 0.3$ unless specified otherwise) are used as the soft and hard magnetic layers, respectively, due to the lower (higher) coercivity of the former (latter). We preferred a Tb-based ferrimagnetic metallic alloy over an elemental transition metal such as Co, since the former exhibits a large and tunable bulk PMA over a broad range of composition, thickness, and temperature, independently of the under/over layer, thus eliminating the laborious engineering of the interfacial PMA in the latter.[32–34] The thickness of the Cu spacer is varied between 1 and 5 nm and Cu is chosen due to its high spin conductivity when interfaced with FMGs and TbCo.[35,36] Finally, Pt is used as capping to avoid oxidation of TbCo and it has no active role in the spin transport experiments reported in this paper, therefore will not be mentioned in the remainder of the text.

The magnetization behavior of TbIG and TbCo were examined by polar magneto-optic Kerr effect (MOKE, Fig. 1c) and AHE (Fig. 1d) measurements with a swept out-of-plane field ($H_z$). In the former, we observed a clear hysteresis loop typical of PMA systems with 100% remanence, and large coercivity $H_c \sim 35$ mT. A closer look-up to the lower field region reveals minor signal jumps at $\sim \pm 10$ mT. The inner loop measurement focusing on this region (see inset – dark green curve) indicates a second magnetic signal with a much lower amplitude, superimposed to the larger hysteresis loop. The $H_c$ value of this inner loop corresponds to the one measured for TbIG prior to the Cu|TbCo deposition (see inset - light green curve), hence this signal corresponds to TbIG. The similar coercivity of the TbIG before and after deposition of the Cu|TbCo indicates negligible exchange coupling between the two magnetic layers (see Supplementary Information). The Hall resistance ($R_H$) measurement shows the hysteresis loop originating from the AHE in TbCo with $H_c \sim 35$ mT, consistent with the MOKE result. However, an inner loop measurement similar to Fig. 1c - inset does not show any signal related to the TbIG reversal. Because of the absence of current flow through TbIG, the AHE signal cannot be generated in this layer and the interfacial AHE is also inoperative due to the negligible spin Hall effect in Cu. These measurements clearly demonstrate that the magnetization vector of TbIG in the above stack can only be probed optically but not electrically with standard Hall measurements.

**Magnetoresistive detection of perpendicular switching**

We then examine the longitudinal resistance behavior of the above stack and compare it with reference material systems. The GMR in magnetic transition-metal multilayers is known to originate from the spin-dependent scattering in the spin-split *d* bands,[37,38] typically leading to an increase of the resistance when the magnetization (*M*) of consecutive layers is antiparallel compared to the parallel case. Figure 2a shows the resistance behavior of a fully metallic Co|Cu|TbCo PMA spin valve device during an $H_z$ sweep. We observe two reversal events at low and high fields, corresponding to the switching of Co and TbCo, respectively, resulting in two distinct resistance levels in the parallel and antiparallel configurations, expectedly of the CIP-GMR. The magnitude of the GMR is 4.9%, comparable to previous reports on similar systems.[36,39] Next, we measure Cu/TbCo deposited on a non-magnetic gadolinium gallium garnet ($Gd_3Ga_5O_{12}$, GGG) substrate (Fig. 2b). Here, the spin valve behavior is not observed. Instead, the resistance displays a field-dependent variation resembling a bowtie shape. This unconventional resistance behavior was previously reported and attributed to the magnon magnetoresistance[40] and/or magnetoresistance due to the sperimagnetism of TbCo.[41] This effect requires a conducting magnetic layer and exhibits a quasi-linear positive or negative slope depending on the relative direction of *M* and $H_z$.

Figure 2c shows the magnetoresistance of the TbIG|Cu|TbCo trilayer. The resistance clearly undergoes four distinct switching events corresponding to the perpendicular switching of TbIG and TbCo, superimposed to the quasi-linear background magnetoresistance described above. Initially, at large positive fields, the magnetizations of TbIG and TbCo are parallel and aligned with $H_z$. Upon reducing $H_z$, the resistance increases linearly until ~-10 mT at which field the magnetization of TbIG reverses from up to down and a sudden resistance increase occurs. As $H_z$ is reduced further, the

resistance increase continues, until the coercivity of TbCo is reached at ~-35 mT, where the resistance drops sharply, followed by further gradual $H_z$-dependent decrease. A reciprocal sequence is observed when $H_z$ is swept from the negative to positive direction. To better illustrate the GMR-like behavior, in Fig. 2d, we plot the magnetoresistance after subtraction of the background due to the unusual magnetoresistance of TbCo not related to the GMR effect. Finally, Fig. 2e shows the minor loops, corresponding to the reversal of $M$ of TbIG, when the TbCo layer is pre-set in the up (+z) and down (-z) direction. In these measurements, we observe the hysteretic resistance behavior in both cases with an opposite sign due to the reversal of the reference layer. Based on the amplitude of the jump in Fig. 2e, we estimate the relative change of resistance $\Delta R/R_0 = 1.4 \times 10^{-5}$, where $R_0$ is the resistance at zero applied field.

The signals reported in Fig. 2c-e are consistent with a positive GMR in TbIG|Cu|TbCo. The sign of the GMR in all-metallic systems is dictated by the spin-dependent scattering asymmetry at the bulk and interfaces of the magnetic layers. In TbCo, the spin-polarized conduction is mostly dominated by the *s-d* electrons of Co, hence the Co magnetization dictates the bulk and interface contribution to the GMR as well as the sign of the AHE. The positive AHE in Fig. 1d and the positive GMR in Fig. 2a establish that the net magnetization in TbCo is parallel to the Co sublattice in the TbIG|Cu|TbCo stack. Likewise, independent measurements reported in our earlier study (see Ref.[12]) and Supplementary Information show that the net magnetization in TbIG is Fe sublattice dominated. Since TbIG is an insulator, only the TbIG/Cu interface contributes to the GMR. Therefore, the positive GMR in TbIG|Cu|TbCo demonstrates that the sign of the spin-dependent scattering asymmetry at both the TbIG|Cu and Cu|TbCo interfaces is the same and positive with respect to the standard convention.

**Temperature-dependence of the magnetoresistance**

To examine further the effect of the relative orientation of the ferrimagnetic sublattices on the sign, and the influence of the measurement temperature on the amplitude of $\Delta R/R_0$, we studied the magnetoresistance in TbIG|Cu|TbCo as a function of temperature ($T$), across the magnetic compensation temperature ($T_M$) of TbIG. In Fig. 3a, we plot representative data during a $H_z$ sweep measured at $T = 225$ K and 175 K, i.e., above and below the $T_M$ of TbIG, respectively. We found $T_M$ of TbIG ~200 K for this specific sample (see Supplementary Information). Because of the linear magnetoresistance background from TbCo the overall shape of the signal becomes intricate for $T \leq 200$ K. Nevertheless, we observe that the resistance changes due to TbIG reversal (marked with black arrows) is upward for $T = 225$ K and downward for $T = 175$ K, indicating a clear sign change of $\Delta R/R_0$. This is further highlighted in Fig. 3b with the minor loop measurements corresponding to the magnetization reversal of TbIG when the TbCo layer is fixed along the up (+z) direction. Above $T_M$ ($T = 225$ K) we observe a negative hysteresis loop (similar to the room temperature measurement reported in Fig. 2e) whereas below $T_M$ ($T = 175$ K) the loop becomes positive.

In Fig. 3c we plot $\Delta R/R_0$ in the entire temperature range tested in this study. The sign reversal of $\Delta R/R_0$ at $T_M$ of TbIG confirms that the spin-dependent reflection at the TbIG|Cu interface, similarly to the interfacial AHE[42], is not governed by the net magnetization, but instead, by the orientation of the Fe sublattice. In spite of the sign change at $T \sim 200$ K, the absolute value of $\Delta R/R_0$ increases smoothly and monotonically (open dots in Fig. 3c) with the decreasing temperature. This is expected from the strong temperature dependence of the GMR effect due to the increased spin conductivity of Cu at lower temperatures.[43]

**TbCo composition and Cu spacer thickness dependence**

The amplitude and sign of the GMR can be modulated by the TbCo composition and Cu spacer thickness. To investigate the former, we measured $\Delta R/R_0$ in TbIG|Cu(2)|Tb$_x$Co$_{1-x}$(8nm) layers with $x$ ranging between 0.3 and 0.65 and plotted in Fig. 4 – left. We note that for $x < 0.3$ the films were magnetized in-plane and for $x > 0.65$ they were paramagnetic (see Supplementary Information). We find that $\Delta R/R_0$ strongly depends on the Tb content, displaying a rapid decrease going from $x = 0.3$ to 0.5. When the Tb amount reaches $x = 0.55$, the magnetization becomes Tb sublattice dominated and $ΔR/R_0$ becomes negative (i.e. $R$ is lower when the magnetization of TbIG and $Tb_xCo_{1-x}$ are antiparallel). This is analogous to the sign reversal observed upon crossing the $T_M$ of TbIG reported in Fig. 3c, but obtained at room temperature with a modification of the $Tb_xCo_{1-x}$ composition instead. Overall, $|ΔR/R_0|$ undergoes a ~22-fold decrease when the Tb concentration is increased from 0.3 to 0.6, and finally vanishes at $x = 0.65$. This result indicates the essential role played by Co in the spin-polarized current generation and spin-dependent scattering in the bulk and interfaces of TbCo.

Figure 4 – right reports the effect of the Cu spacer thickness ($t$) on $ΔR/R_0$ in TbIG|Cu($t$)|TbCo layers. The largest effect is found for Cu(2 nm). Increasing $t$ results in a decrease of $ΔR/R_0$, as expected due to the increased current shunting and reduced spin coherence.[43] On the other hand, for a Cu thickness of 1 nm, a slightly lower $ΔR/R_0$ is observed with respect to Cu(2 nm), whereas the effect is, in principle, expected to be enhanced. We associate this behavior to the lower uniformity of such thin Cu layer, causing additional resistance, not participating in the magnetoresistance and negatively affecting the spin-dependent properties and interface quality with the TbCo layer. Indeed, we observe a correlation between the Cu thickness and the coercivity and PMA of TbCo, indicating a sharper interface when the Cu thickness is increased, promoting a higher quality TbCo. This might be the reason why the magnetoresistance is still substantial despite larger current shunting when the Cu spacer is increased to 5 nm. The strong dependence of the magnetic properties of TbCo alloys on the underlayer material, thickness and interface quality is well known.[44] Additionally, discrepancies in the spin-dependent properties of the TbIG|Cu interface for TbIG grown on different dates on different substrates could influence the data presented in Fig. 4 – right. Although the essential features and trends expected of the thickness dependence of $ΔR/R_0$ are observed, the data in Fig. 4 – right cannot reflect a universal behavior. Therefore, we believe that a more systematic study of the influence of the spacer layer thickness on $ΔR/R_0$ should be conducted by continuously varying the Cu thickness on a single substrate to ensure the uniformity of the TbIG properties, which falls outside our technical capabilities. Finally, to confirm the relevance of the spacer layer with low spin-orbit coupling (high

spin conductivity) on the magnetoresistance, we replaced Cu with 2 nm of Pt which exhibits large spin-orbit coupling. We did not detect any spin valve effect within our experimental detection limit. This negative result is anticipated due to the very short spin diffusion length of Pt (typically <2 nm), dephasing nearly the entire spin current generated in the TbCo and at the Pt/TbCo interface.[36] Furthermore, the absence of spin valve signal with the Pt spacer excludes the magnetic proximity effect at the TbIG|metal interface as the potential origin of the magnetoresistance, since Pt is much more susceptible to proximity magnetism than Cu. We note that $\Delta R/R_0$ values of all the materials examined in this work can be found in Table S1 of the Supplementary Information.

**Boltzmann Transport Calculations**

We calculated the magnetization-dependent resistivity of TbIG|Cu|TbCo trilayer by solving the following linearized Boltzmann transport equation with a layer-by-layer approach:[38,45]

$$v_z \frac{\partial g}{\partial z} - \frac{eE}{m}\frac{\partial f_0}{\partial v_x} = -\frac{g}{\tau} \quad (1)$$

Here $m$ and $\tau$ are the effective mass and momentum relaxation time and mass of conduction electrons, $f_0$ is the equilibrium electron distribution function, $g(\boldsymbol{v}, z)$ is the deviation from $f_0$ induced by the external electric field with amplitude $E$ applied along **x**. Full translation symmetry is assumed in the x-y plane (i.e., the layer plane) so that the spatial dependence of $g$ only occurs in the thickness direction parallel to the z-axis. To further simplify our calculations, we assumed the same effective mass, relaxation time, and Fermi energy $\varepsilon_F$ for the conducting Cu and TbCo layers, which would be sufficient for the purpose of an order of magnitude estimation of the $\Delta R/R_0$ ratio of the trilayer system. The calculation can be extended straightforwardly to involve different values of the above-mentioned materials parameters without essential change to the formulation.

In order to capture the interfacial spin-dependent scattering, we divide $g$ into four additive components depending on the orientation of an electron's spin moment with respective to the magnetization of a magnetic layer and the sign of $v_z$ (i.e., the z-component of the electronic group velocity). The general solutions of Eq. (1) can be written as

$$g_{\pm\uparrow(\downarrow)}(\boldsymbol{v},z) = \frac{eE\tau}{m}\frac{\partial f_0}{\partial v_x}\left[1 + C_{\pm\uparrow(\downarrow)}exp\left(\frac{\mp z}{\tau|v_z|}\right)\right], \qquad (2)$$

where the "+" and "−" signs denote electrons with $v_z$ being positive and negative, respectively, the arrow ↑ (↓) characterizes orientation of spin parallel (antiparallel) to the local magnetization of the magnetic layer in question. The general solutions take the same form for electrons in the Cu and TbCo layers, except for the coefficients, $C_{\pm\uparrow(\downarrow)}$, which need to be determined by boundary conditions.

To calculate the $\Delta R/R_0$ ratio, scatterings at three interfaces of the trilayer need to be taken into account through the boundary conditions, namely, 1) the spin-dependent reflection at the out surface of the TbCo layer, 2) the specular transmission and reflection at the interface between the metallic TbCo|Cu interface, and 3) the spin-dependent specular reflection at the Cu|TbIG interface. In addition, when the magnetizations of the TbCo and TbIG layers are antiparallel, we also take into account the change of spin quantization axis in the middle of the Cu layer. It is reasonable to assume specular reflection at the metallic interface is negligible and scattering at the outer surface of the TbCo layer is completely diffusive. And near the magnetic interface between the Cu and TbIG layer, the portions of electron fluxes that are specularly reflected from the interface are, in principle, also spin dependent,[21] as "spin-up" and "spin-down" electrons see different energy barriers effectively due to the exchange coupling ($J_{ex}$) between them and the magnetization of the TbIG layer even though the magnetic layer is insulating.

By inserting the general solutions of $g_{\pm\uparrow(\downarrow)}$ for each layer into these boundary conditions, the unknowns in Eqs. (2) can be determined, allowing us to further evaluate the spatially-averaged longitudinal conductivity of the $i^{th}$ layer via $\sigma^{(i)} = \frac{1}{d_i E}\int dz \int d^3\boldsymbol{v}\, v_x(g_\uparrow + g_\downarrow)$, where $d_i$ is the thickness of the $i^{th}$ layer. The total resistivities for parallel and antiparallel magnetization configurations, denoted by $\rho_{\uparrow\uparrow}$ and $\rho_{\uparrow\downarrow}$ respectively, are obtained by inverting the corresponding conductivity tensors. Finally, the $\mathit{\Delta R}/R_0$ ratio is evaluated via $\Delta R/R_0 = (\rho_{\uparrow\downarrow} - \rho_{\uparrow\uparrow})/\rho_{\uparrow\uparrow}$. For the TbCo|Cu interface, the transmission coefficients are taken to be $T_\uparrow = 0.5$ and $T_\downarrow = 0.95$ (note that the roughness of the interface can be characterized phenomenologically by the spin-dependent diffusive scattering parameter defined as $D_{\uparrow(\downarrow)} = 1 - T_{\uparrow(\downarrow)}$.[45] For $J_{ex} \sim 0.01$ eV and the averaged energy barrier of the insulator $V_b \sim 12$ eV,[2,21] the reflection coefficients are estimated to be $R_\uparrow = 0.4995$ and $R_\downarrow = 0.5005$. This spin asymmetry of electron scatterings at the two magnetic interfaces gives rise to a $\mathit{\Delta R}/R_0$ ratio of 6.2x10$^{-5}$, close to the experimental value of 7.0x10$^{-5}$ measured at 10 K. Detailed calculations are provided in the Supplementary Information.

**Conclusions**

In summary, we demonstrate a simple magnetoresistive detection of perpendicular magnetization reversal in an insulating ferrimagnetic material TbIG. The detection relies on current-in-plane magnetoresistance measurements in a TbIG|Cu|TbCo trilayer system where the conducting ferrimagnet TbCo is used as a reference magnetic layer and spin polarizer. The material and temperature dependence of the magnetoresistance and theoretical calculations collectively pinpoint the spin-valve effect at the TbIG|Cu interface as the underlying cause of the observed phenomenon. These results will open a new chapter in the field of insulating spintronics as they will stimulate research into a wide spectrum of FMG, spacer and spin polarizer layer combinations, and enable a whole new range of device ideas and architectures based on magnetic insulators. While the effect is relatively small for any microelectronic applications as of yet, it can be enhanced by orders of magnitude by materials and device engineering, and support from theory.

On the materials side, ferromagnetic materials with higher spin polarization such as Heusler alloys with an optimized thickness could provide a pathway to increase the magnetoresistance. To obtain even larger gains, the spacer layer can be chosen from those having a very long spin coherence length and promoting better spin mixing conductance with the insulating and conducting magnetic layers forming the device. Potential candidates include two-dimensional materials such as graphene, transition metal dichalcogenides (TiSe$_2$, MoS$_2$, etc.), conducting oxides (SrVO$_3$, etc.) and other light metals (Cr, Mn, etc.). Our work will also stimulate theoretical efforts, as the most suitable materials could be determined by using relevant first-principles calculations.

On the device side, the magnetoresistive reading reported here will enable non-volatile binary memory cells where a magnetic insulator could be used as an active component instead of conventional magnetic conductors. In such devices, using a magnetic insulator would bring about a series of advantages such as higher structural stability, broader magnetic tunability, ultrafast switching times, and low power consumption, among others. It is also possible to use the magnetoresistance output to identify and characterize the skyrmions and domain walls in an insulating racetrack memory and study their field/current-driven dynamics in real time. Insulating domain wall- and skyrmion-based devices enabled by magnetoresistive reading could pave the way for novel analog-like memory concepts that can be used in neuromorphic computing.

**Methods**

**Samples preparation**

25 nm-thick TbIG thin films were deposited on GGG(111) substrates by radio frequency (r.f.) sputtering at 800 ºC from a stoichiometric target. The deposition rate was ~0.4 nm/min at the applied power of 150 W in 3 mTorr of a mixture of Ar and O$_2$ with a ratio of 30:2 and the base pressure in the chamber was below 7x10$^{-8}$ Torr. Detailed characterization and optimization procedures of our TbIG films are described in Ref.[12]. To fabricate the six-terminal Hall bar devices, the continuous TbIG films were covered in photoresist and patterned by laser-writer optical lithography. Finally, the metallic stack described further below was deposited and lift-off was performed. The Hall bar dimensions are *l*=30 μm for the current line length, *l*/4 its width and *l*/10 the Hall branch width.

The metallic stack consisted of M(*t*)|Tb$_x$Co$_{1-x}$(8 nm)|Pt(3 nm) multilayers deposited by d.c. magnetron sputtering at room temperature in 3 mTorr Ar. The Tb$_x$Co$_{1-x}$ alloys were obtained by co-sputtering

pure Co and Tb targets and the relative atomic concentration of the two elements was controlled by the relative sputtering power. The thickness *t* of the metal (M) spacer layer varied between 1 and 5 nm. The deposition rates were as follows: 0.13514 nm/s at 100 W for Cu, 0.104 nm/s at 200 W for Co, 0.082 nm/s t 50 W for Tb, 0.186 nm/s at 50 W for Pt and 0.057 nm/s at 200 W for Ti. The following reference samples were also prepared: Cu(2 nm)|$Tb_{0.3}Co_{0.7}$(8 nm)|Pt(3 nm) on GGG(111) substrate, Ti(3 nm)|Pt(3 nm)|Co(1 nm)|Cu(2 nm)|$Tb_{0.3}Co_{0.7}$(8 nm)|Pt(3 nm) on Si, Ti(3 nm)|Pt(3 nm)|Co(1 nm)|Cu(2 nm)|$Tb_{0.65}Co_{0.35}$(8 nm)|Pt(3 nm) on Si and Pt(1.5 nm)|Cu(1.5 nm)|$Tb_{0.3}Co_{0.7}$(8 nm)|Pt(3 nm) on TbIG.

**Magnetic and Electrical Characterization**

The magnetic hysteresis and anisotropy axis of the films were examined using a home-built magneto-optic Kerr effect (MOKE) setup in a polar geometry with a 532 nm wavelength green laser. The room temperature magnetoresistance measurements were performed by recording the resistance (*R*) as a function of *H* between the extremes of the current line using a Keithley DMM6500 digital multimeter and DC test current of 1 mA. For the temperature dependence of the magnetoresistance, selected samples were inserted inside a physical property measurement system and measured in the range of 10–300 K with an AC probing current of 1 mA (root mean square) and frequency $\omega/2\pi = 999$ Hz. The Hall effect measurements were performed at room temperature using a Zurich Instruments MFLI digital lock-in amplifier. An AC probing current of amplitude 0.5-1 mA (root mean square) and frequency $\omega/2\pi = 1092$ Hz was sent across the current line and the first harmonic voltage ($R_H$) was measured across the Hall arms.

**Data availability**

All the data supporting the findings of this study are available upon request from the corresponding author.

**Code availability**

The computer code used for data analysis is available upon request from the corresponding author.

**References**

1. Yang, Y., Liu, T., Bi, L. & Deng, L. Recent advances in development of magnetic garnet thin films for applications in spintronics and photonics. *J. Alloys Compd.* **860**, 158235 (2021).

2. Kajiwara, Y. *et al.* Transmission of electrical signals by spin-wave interconversion in a magnetic insulator. *Nature* **464**, 262–266 (2010).

3. Uchida, K. I. *et al.* Observation of longitudinal spin-Seebeck effect in magnetic insulators. *Appl. Phys. Lett.* **97**, 172505 (2010).

4. Nakayama, H. *et al.* Spin Hall Magnetoresistance Induced by a Nonequilibrium Proximity Effect. *Phys. Rev. Lett.* **110**, 206601 (2013).

5. Cornelissen, L. J., Liu, J., Duine, R. A., Youssef, J. Ben & Van Wees, B. J. Long-distance transport of magnon spin information in a magnetic insulator at room temperature. *Nat. Phys. 2015 1112* **11**, 1022–1026 (2015).

6. Meyer, S. *et al.* Observation of the spin Nernst effect. *Nat. Mater.* **16**, 977–981 (2017).

7. Seifert, T. S. *et al.* Femtosecond formation dynamics of the spin Seebeck effect revealed by terahertz spectroscopy. *Nat. Commun.* **9**, 2899 (2018).

8. Caretta, L. *et al.* Relativistic kinematics of a magnetic soliton. *Science* **370**, 1438–1442 (2020).

9. Chumak, A. V., Vasyuchka, V. I., Serga, A. A. & Hillebrands, B. Magnon spintronics. *Nat. Phys.* **11**, 453–461 (2015).

10. Avci, C. O. *et al.* Current-induced switching in a magnetic insulator. *Nat. Mater.* **16**, 309–314 (2017).

11. Quindeau, A. *et al.* Tm3Fe5O12/Pt Heterostructures with Perpendicular Magnetic Anisotropy for Spintronic Applications. *Adv. Electron. Mater.* **3**, 1600376 (2017).

12. Damerio, S. & Avci, C. O. Sputtered terbium iron garnet films with perpendicular magnetic anisotropy for spintronic applications. *J. Appl. Phys.* **133**, 073902 (2023).

13. Shao, Q. *et al.* Role of dimensional crossover on spin-orbit torque efficiency in magnetic insulator thin films. *Nat. Commun.* **9**, 3612 (2018).

14. Avci, C. O. *et al.* Interface-driven chiral magnetism and current-driven domain walls in insulating magnetic garnets. *Nat. Nanotechnol.* **14**, 561–566 (2019).

15. Vélez, S. *et al.* High-speed domain wall racetracks in a magnetic insulator. *Nat. Commun.* **10**, 4750 (2019).

16. Avci, C. O. Current-induced magnetization control in insulating ferrimagnetic garnets. *J. Phys. Soc. Japan* **90**, 081007 (2021).

17. Vélez, S. *et al.* Current-driven dynamics and ratchet effect of skyrmion bubbles in a ferrimagnetic insulator. *Nat. Nanotechnol.* **17**, 834–841 (2022).

18. Avci, C. O. *et al.* Nonlocal Detection of Out-of-Plane Magnetization in a Magnetic Insulator by Thermal Spin Drag. *Phys. Rev. Lett.* **124**, 027701 (2020).

19. Huang, S. Y. *et al.* Transport magnetic proximity effects in platinum. *Phys. Rev. Lett.* **109**, 107204 (2012).

20. Chen, Y. T. *et al.* Theory of spin Hall magnetoresistance. *Phys. Rev. B* **87**, 144411 (2013).

21. Zhang, S. S. L. & Vignale, G. Nonlocal Anomalous Hall Effect. *Phys. Rev. Lett.* **116**, 136601 (2016).

22. Meyer, S. *et al.* Anomalous Hall effect in YIG|Pt bilayers. *Appl. Phys. Lett.* **106**, 132402 (2015).

23. Baibich, M. N. *et al.* Giant magnetoresistance of (001)Fe/(001)Cr magnetic superlattices. *Phys. Rev. Lett.* **61**, 2472–2475 (1988).

24. Binasch, G., Grünberg, P., Saurenbach, F. & Zinn, W. Enhanced magnetoresistance in layered magnetic structures with antiferromagnetic interlayer exchange. *Phys. Rev. B* **39**, 4828 (1989).

25. Yuasa, S., Nagahama, T., Fukushima, A., Suzuki, Y. & Ando, K. Giant room-temperature magnetoresistance in single-crystal Fe/MgO/Fe magnetic tunnel junctions. *Nat. Mater.* **3**, 868–871 (2004).

26. Parkin, S. S. P. *et al.* Giant tunnelling magnetoresistance at room temperature with MgO (100) tunnel barriers. *Nat. Mater.* **3**, 862–867 (2004).

27. Daughton, J. M. GMR applications. *J. Magn. Magn. Mater.* **192**, 334–342 (1999).

28. B. Dieny. Giant magnetoresistance in spin-valve multilayers. *J. Magn. Magn. Mater.* **136**, 335–359 (1994).

29. Snoeck, E. *et al.* Experimental evidence of the spin dependence of electron reflections in magnetic CoFe2O4/Au/Fe3O4 trilayers. *Phys. Rev. B* **73**, 104434 (2006).

30. Van Dijken, S., Fain, X., Watts, S. M. & Coey, J. M. D. Negative magnetoresistance in Fe3O4/Au/Fe spin valves. *Phys. Rev. B* **70**, 052409 (2004).

31. Tripathy, D. & Adeyeye, A. O. Giant magnetoresistance in half metallic Fe3O4 based spin valve structures. *J. Appl. Phys.* **101**, 09J505 (2007).

32. Lee, J. W., Park, J. Y., Yuk, J. M. & Park, B. G. Spin-Orbit Torque in a Perpendicularly Magnetized Ferrimagnetic Tb - Co Single Layer. *Phys. Rev. Appl.* **13**, 044030 (2020).

33. Finley, J. & Liu, L. Spin-Orbit-Torque Efficiency in Compensated Ferrimagnetic Cobalt-Terbium Alloys. *Phys. Rev. Appl.* **6**, 054001 (2016).

34. Hansen, P., Klahn, S., Clausen, C., Much, G. & Witter, K. Magnetic and magneto-optical properties of rare-earth transition-metal alloys containing Dy, Ho, Fe, Co. *J. Appl. Phys.* **69**, 3194–3207 (1991).

35. Du, C., Wang, H., Yang, F. & Hammel, P. C. Enhancement of Pure Spin Currents in Spin Pumping Y3Fe5O12/Cu/Metal Trilayers through Spin Conductance Matching. *Phys. Rev. Appl.* **1**, 044004 (2014).

36. Avci, C. O., Lambert, C. H., Sala, G. & Gambardella, P. A two-terminal spin valve device controlled by spin-orbit torques with enhanced giant magnetoresistance. *Appl. Phys. Lett.*

**119**, 032406 (2021).

37. Parkin, S. S. P. Origin of enhanced magnetoresistance of magnetic multilayers: Spin-dependent scattering from magnetic interface states. *Phys. Rev. Lett.* **71**, 1641–1644 (1993).

38. Hood, R. Q. & Falicov, L. M. Boltzmann-equation approach to the negative magnetoresistance of ferromagnetic-normal-metal multilayers. *Phys. Rev. B* **46**, 8287–8296 (1992).

39. Gottwald, M. *et al.* Magnetoresistive effects in perpendicularly magnetized Tb-Co alloy based thin films and spin valves. *J. Appl. Phys.* **111**, 083904 (2012).

40. Mihai, A. P., Attané, J. P., Marty, A., Warin, P. & Samson, Y. Electron-magnon diffusion and magnetization reversal detection in FePt thin films. *Phys. Rev. B* **77**, 060401(R) (2008).

41. Park, J. *et al.* Unconventional magnetoresistance induced by sperimagnetism in GdFeCo. *Phys. Rev. B* **103**, 014421 (2021).

42. Rosenberg, E. R. *et al.* Magnetism and spin transport in rare-earth-rich epitaxial terbium and europium iron garnet films. *Phys. Rev. Mater.* **2**, 094405 (2018).

43. Tripathy, D., Adeyeye, A. O. & Shannigrahi, S. Effect of spacer layer thickness on the magnetic and magnetotransport properties of Fe3O4/Cu/Ni80Fe20 spin valve structures. *Phys. Rev. B* **75**, 022403 (2007).

44. Tolley, R. *et al.* Generation and manipulation of domain walls using a thermal gradient in a ferrimagnetic TbCo wire. *Appl. Phys. Lett.* **106**, 242403 (2015).

45. Camley, R. E. & Barna, J. Theory of giant magnetoresistance effects in magnetic layered structures with antiferromagnetic coupling. *Phys. Rev. Lett.* **63**, 664 (1989).

**Acknowledgments**

S.D. and C.O.A. acknowledge funding from the European Research Council (ERC) under the European Union's Horizon 2020 research and innovation program (project MAGNEPIC, grant agreement No. 949052). C. O. A. acknowledges funding from the Spanish Ministry of Science and Innovation through grant reference no. CNS2022-136060. Work by A.S., M.M. and S.S.-L.Z was supported by the College of Arts and Sciences, Case Western Reserve University. Authors thank M. Fettizio for the insightful discussions.

**Author Information**

**Contributions**

C.O.A. conceived the idea and supervised the study. S.D. designed and prepared the samples, carried out the measurements and analyzed the data. A.S., M.M. and S.S.-L.Z. constructed the theoretical framework. S.D. and C.O.A. wrote the manuscript. All authors discussed the results and commented on the manuscript.

**Corresponding author**

Correspondence to Silvia Damerio (sdamerio@icmab.es) and Can Onur Avci (cavci@icmab.es)

**Competing Interests**

The authors declare no competing interests.

**Figures and Tables**

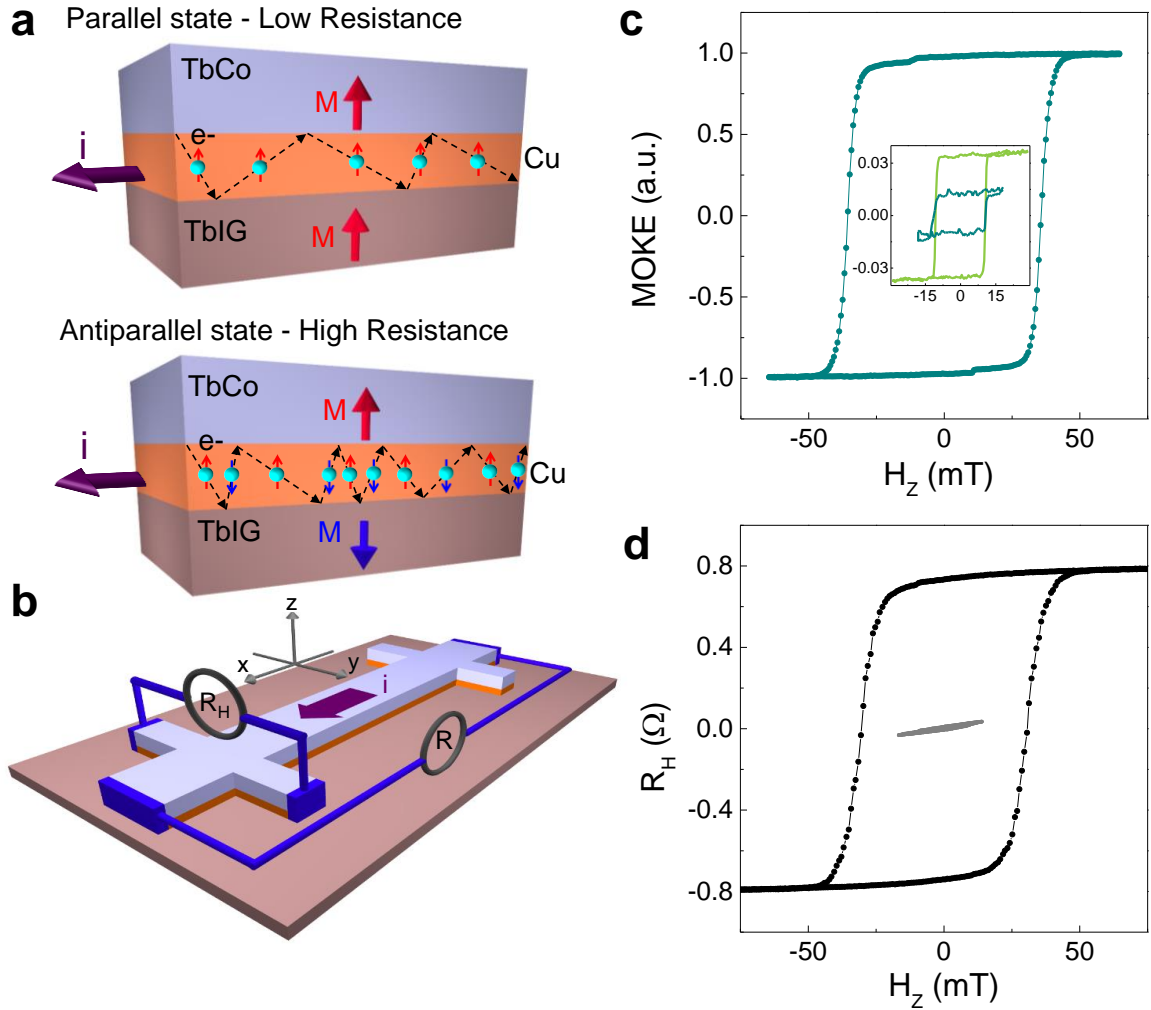

**Fig.1 Illustration of the spin valve effect in TbIG|Cu|TbCo, device schematics, and optical and electrical characterization. a** Schematic representation of the spin valve structure in the high and low resistance states at room-temperature. The red and blue arrows indicate the direction of the magnetization (*M*) and corresponding spins of the majority and minority carriers. Cyan spheres represent conduction electrons undergoing lower and higher scattering events in the parallel and antiparallel magnetic configurations, respectively. **b** Schematic representation of the device with the geometry used for the electrical measurements. **c** Plot of the polar MOKE signal of TbIG|Cu|TbCo. The inset shows the minor loops corresponding to the switching of TbIG prior (light green) and after (dark green) the deposition of TbCo. **d** Transverse Hall resistance ($R_H$) in TbIG|Cu|TbCo as a function

of out-of-plane field ($H_Z$), corresponding to the AHE in TbCo. No fingerprints of TbIG switching are observed in the inner loop (gray).

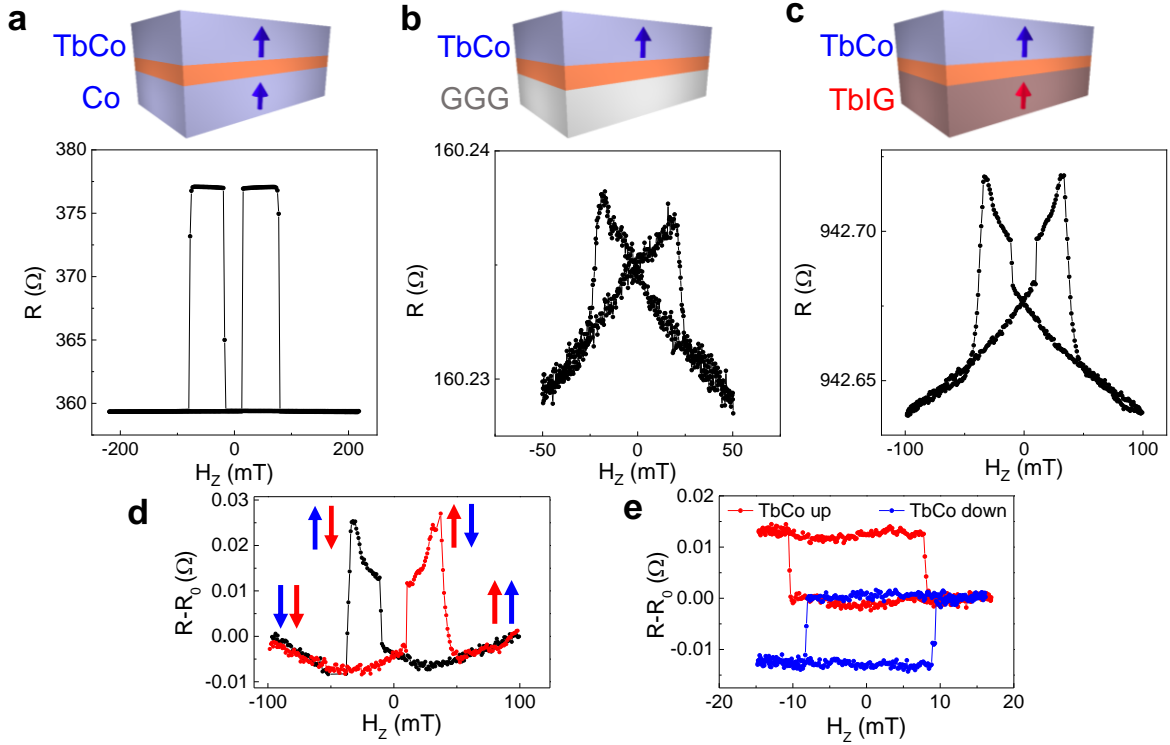

**Fig.2 Magnetoresistance in TbIG|Cu|TbCo and other reference layers. a** Resistance ($R$) as a function of the out-of-plane field ($H_Z$) in an all-metallic Co(1)|Cu(2)|TbCo(8) trilayer. **b** in GGG(subs.)|Cu(2)|TbCo(8) and **c** in TbIG(25)|Cu(2)|TbCo(8). The latter clearly shows a spin valve signal similar to the one observed in the all-metallic trilayer. **d** Plot of the resistance from panel c after removal of the unconventional magnetoresistance contribution from TbCo (see text for more details). The red and blue arrows indicate the direction of *M* in TbIG and TbCo, respectively. **e** Inner loops showing the switching of TbIG when the direction of TbCo is fixed up (red) or down (blue) with respect to the film normal.

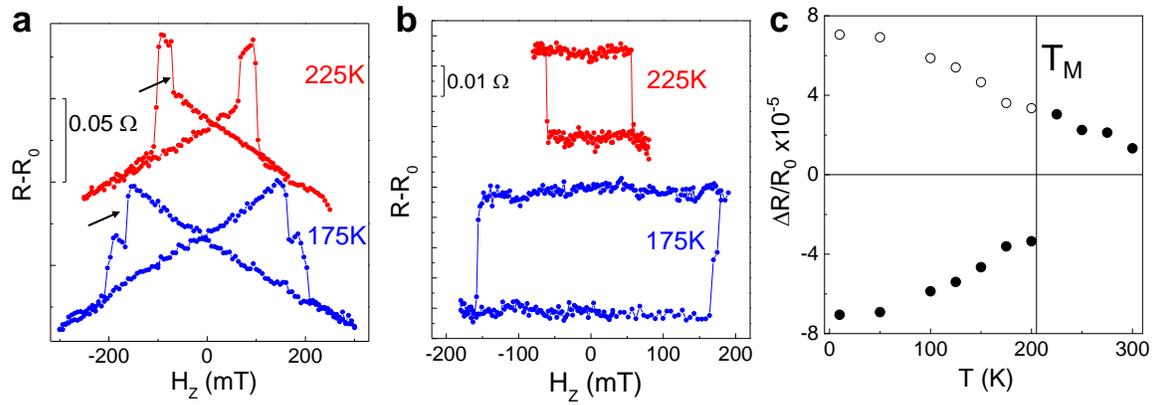

**Fig.3 Temperature dependence of the magnetoresistance in TbIG|Cu|TbCo. a** Plot of the resistance (*R*) as a function of out-of-plane field ($H_Z$) for 2 selected temperatures above (red) and below (blue) the compensation temperature ($T_M$) of TbIG. The black arrows indicate the resistance jump corresponding to TbIG magnetization reversal. **b** Plot of the inner loops showing the switching of TbIG when *M* of TbCo is fixed out-of-plane for two selected temperatures above (red) and below (blue) $T_M$. The linear background has been subtracted. **c** Plot of the temperature (*T*) dependence of the magnetoresistance (close dots) and its absolute value (open dots).

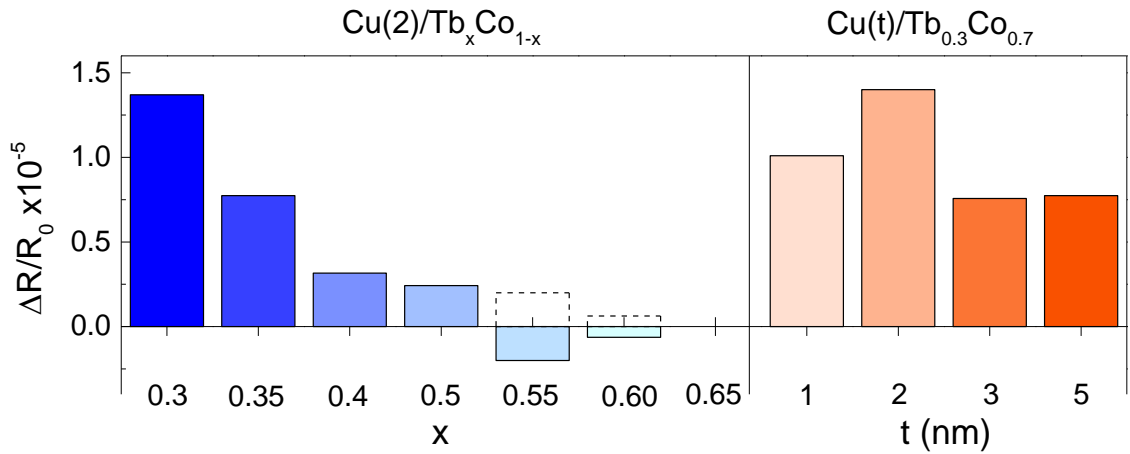

**Fig.4 Amplitude of the magnetoresistance as a function of $Tb_xCo_{1-x}$ composition and Cu spacer thickness**. Left: plot of $\Delta R/R_0$ in TbIG(25)|Cu(2)|$Tb_xCo_{1-x}$(8) alloy as a function of Tb content (*x*) from 0.3 to 0.65. We note that $\Delta R/R_0$ undergoes a sign change for *x* > 0.5 and it is below our detection

limit for $x = 0.65$. The dotted lines represent the absolute value of the GMR amplitude for visual illustration of the monotonic decay of $\Delta R/R_0$. Right: plot of $\Delta R/R_0$ in TbIG(25)/Cu(t)/Tb$_{0.3}$Co$_{0.7}$(8) as a function of Cu spacer thickness ($t$) from 1 to 5 nm.

# Supplementary Information for "Magnetoresistive detection of perpendicular switching in a magnetic insulator"

Silvia Damerio[1,*], Achintya Sunil[2], M. Mehraeen[2], Steven S.-L. Zhang[2], and Can O. Avci[1,*]

[1]Institut de Ciència de Materials de Barcelona, Campus de la UAB, Bellaterra, 08193, Spain

[2]Department of Physics, Case Western Reserve University, Cleveland, Ohio 44106, USA

**SI1 – Exchange bias in TbIG|Cu|TbCo spin valves**

We found that the magnetization of the magnetic layers in the spin valves of this study is coupled via a small exchange bias for the Cu thickness ≤2 nm. This interaction appears as a lateral shift towards negative (positive) values of the hystereis loops of the "soft" TbIG when the magnetization of $Tb_{0.3}Co_{0.7}$ is fixed in the up (down) direction (see Fig. 2e of the main text). We estimated the shift ($\Delta H$) as a function of the thickness of the Cu spacer layer and plot the trend in Fig. S1. Due to the low amount of samples available we could not perform a detailed statistical analysis of these values, but the results seem to indicate that the exchange bias decreases upon increasing spacer layer thickness. Because the measured exchange bias is much lower than the coercivity of TbIG, we consider it negligible in this study.

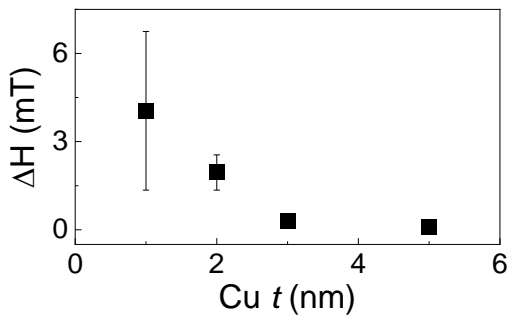

**Fig.S1 Exchange bias in TbIG|Cu|TbCo spin valves. a** Plot of the difference of the coercive field of TbIG ($\Delta H$), measured from the *R* vs *H* loops, when the magnetization of TbCo is fixed in the up and down direction as a function of the Cu spacer layer thickness (*t*).

**SI2 – Magnetic and magneto-transport properties of $Tb_xCo_{1-x}$ alloys**

Ferrimagnets are a class of magnets with unbalanced antiparallel-aligned sublattice moments, which results in a finite, albeit small, magnetization. In RE-transition metal alloys and multilayers, the dominant sublattice (i.e. the sublattice whose moment is parallel to the net magnetization) varies with the stoichiometry and temperature. Measurement of the anomalous Hall effect (AHE) provides information on the dominating sublattice, as the sign of the AHE coefficient is opposite for RE and transition-metals.

Before utilizing $Tb_xCo_{1-x}$ in the fabrication of spin valve devices we characterized the magnetic and magneto-transport properties of the alloy as a function of Tb concentration. To this end, we grew 8 nm thick $Tb_xCo_{1-x}$ films by co-sputtering on Si substrates with 3 nm Ti buffer and capping layers. The samples were patterned into Hall-bar devices (with sizes $l$=30 µm for the current line length, $l/4$ its width and $l/10$ the Hall branch width) by standard optical lithography and lift-off. Figure S2a shows the plot of the Hall resistance ($R_H$) as a function of out-of-plane field ($H_Z$) of the $Tb_xCo_{1-x}$ films below the compensation composition. Here, PMA is only achieved above 25% of Tb due to the shape anisotropy that tends to orient the magnetization in-plane when the films are patterned into micron-sized devices. Here, $H_C$ increases from 65 mT at x=0.25 to 750 mT for x=0.5. Comparable values are observed in the spin valves of the main text (see Table S1) with a small difference due to the different buffer layer (Cu instead of Ti). Between x=0.50 and x=0.55 the compensation composition of the $Tb_xCo_{1-x}$ alloy is reached and the sign of the AHE reverses, as shown in Fig. S2b. Upon further increasing Tb concentration, $H_C$ starts to decrease again, reaching 20 mT at x=0.65. This allows us to conclude that $Tb_xCo_{1-x}$ films are Co-dominated up to x=0.5 and Tb-dominated above x=0.55.

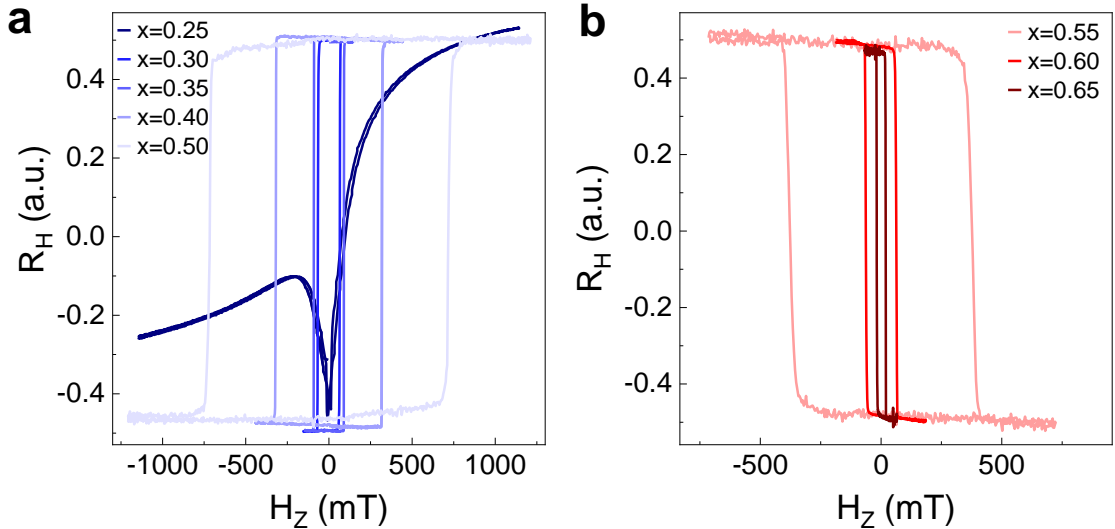

**Fig.S2 Characterization of the AHE in $Tb_xCo_{1-x}$ alloys as a function of Tb content.** Plot of the Hall resistance ($R_H$) as a function of $H_Z$ for patterned $Tb_xCo_{1-x}$ films with **a** x between 0.25 and 0.5 and **b** x between 0.55 and 0.65.

### SI3 – Ferrimagnetic sublattices and sign of the magnetoresistance

The giant magnetoresistance (GMR) provides information on the relative orientation of the moment of the transition metal sublattice in two neighboring layers, as it originates from the scattering of d-conduction electrons with the transition metal moments. Figure S3a shows the AHE of a Co(1)|Cu(2)|$Tb_{0.3}Co_{0.7}$(8) (thickness in nm) reference spin valve. As it can be seen, here the AHE coefficient is positive for both $Tb_{0.3}Co_{0.7}$ and Co layers, indicating that the magnetization ($M$) of $Tb_{0.3}Co_{0.7}$ is parallel to the Co sublattice. Consistently, the GMR (Fig. S3b also Fig. 2a of the main text) is positive, indicating that in both layers the Co sublattice, which dominates the transport properties, is aligned with the magnetic field ($H_Z$). Figures S3c-d show the measurements of a reference Co(1)|Cu(2)|Tb$_{0.65}$Co$_{0.35}$(8) spin valve. Here, due to the large amount of Tb, the alloy becomes RE-dominated and thus both the AHE and the GMR change sign. Figure S3e-f show the Hall resistance ($R_H$) as a function of $H_Z$ in a TbIG(25)|Pt(1.5)|Cu(1.5)|Tb$_{0.3}$Co$_{0.7}$(8) spin valve. In this case, the insertion of a thin (1.5 nm) layer of Pt in contact with the TbIG is necessary to read its magnetization direction via the spin Hall magnetoresistance (SMR) effect. Here, we observe a major hysteresis loop (Fig. S3e) corresponding to the AHE of Tb$_{0.3}$Co$_{0.7}$ with positive AHE coefficient, and a minor loop (Fig. S3f) corresponding to the SMR-AHE of TbIG. The latter is negative, indicating that TbIG is above its compensation temperature and thus its magnetization is parallel to the moment of the tetragonal Fe$^{3+}$ sublattice (notice that 3d metal-dominated TbIG and TbCo have opposite sign of the AHE coefficient). In this type of spin-valve the GMR is positive, as shown in Fig.2c of the main text. On the other hand, Fig. S3g shows the negative AHE in a TbIG(25)|Cu(2)|Tb$_{0.55}$Co$_{0.45}$(8) spin valve, which displays the negative GMR of Fig. S3h, as the TbIG and TbCo are respectively 3d metal and RE dominated. Here, the TbCo coercive field is significantly higher, being closer to the magnetic compensation composition. These data confirm that the transport properties in ferrimagnetic insulator|spacer|metal spin valves are dominated by the transition metal sublattice in both layers and show that the spin-dependent reflection coefficient at the TbIG|Cu interface has the same sign as that of the well-known Co|Cu interface.

**Table S1 Amplitude of the magnetoresistance in different TbIG|spacer|magnetic metal trilayers at room temperature.**

| Metallic layer (t nm) | Spacer layer (t nm) | H$_C$ oxide (mT) | H$_C$ metal (mT) | $\Delta R/R_0$ |
|---|---|---|---|---|
| Tb$_{0.25}$Co$_{0.75}$ (8) | Cu (2) | 7 | In-plane | < detection |
| Tb$_{0.3}$Co$_{0.7}$ (8) | Cu (2) | 10 | 35 | 1.44E-5 |
| Tb$_{0.35}$Co$_{0.65}$ (8) | Cu (2) | 10 | 55 | 7.74E-6 |
| Tb$_{0.4}$Co$_{0.6}$ (8) | Cu (2) | 7 | 150 | 3.16E-6 |
| Tb$_{0.5}$Co$_{0.5}$ (8) | Cu (2) | 6 | 800 | 2.42E-6 |
| Tb$_{0.55}$Co$_{0.45}$ (8) | Cu (2) | 9 | 350 | -2.0E-6 |
| Tb$_{0.6}$Co$_{0.4}$ (8) | Cu (2) | 10 | 65 | -6.3E-7 |
| Tb$_{0.65}$Co$_{0.35}$ (8) | Cu (2) | 6 | 20 | < detection |
| Tb$_{0.3}$Co$_{0.7}$ (8) | Cu (1) | 50 | 0 | 1.01E-5 |
| Tb$_{0.3}$Co$_{0.7}$ (8) | Cu (3) | 10 | 48 | 7.57E-6 |
| Tb$_{0.3}$Co$_{0.7}$ (8) | Cu (5) | 10 | 40 | 7.74E-6 |
| Tb$_{0.3}$Co$_{0.7}$ (8) | Pt (2) | 15 | 73 | < detection |

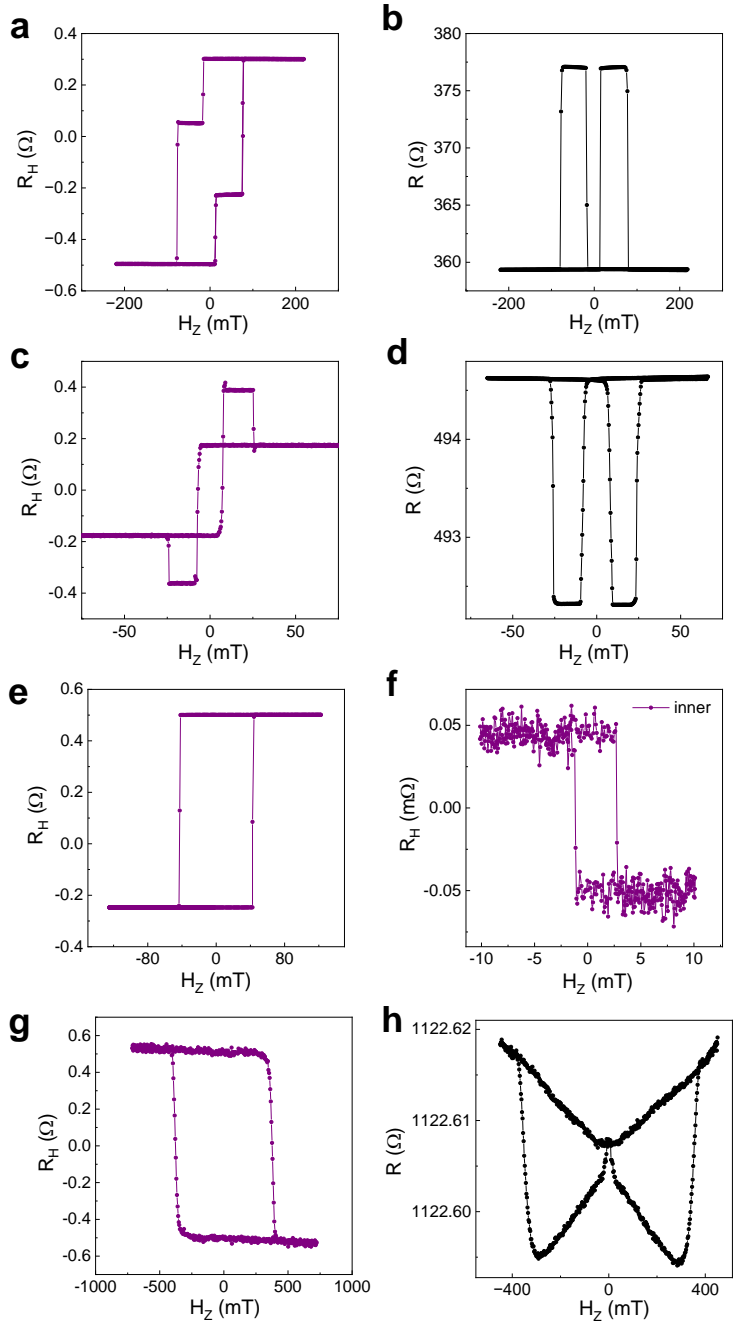

**Fig.S3 AHE and GMR across the ferrimagnetic compensation.** Plot of **a** the Hall resistance ($R_H$) and **b** the longitudinal resistance ($R$) as a function of out-of-plane magnetic field ($H_Z$) of a Co(1)|Cu(2)|Tb$_{0.3}$Co$_{0.7}$(8) reference spin valve. Plot of **c** $R_H$ and **d** $R$ as a function of $H_Z$ of a Co(1)|Cu(2)|Tb$_{0.65}$Co$_{0.35}$(8) reference spin valve. **e-f** Plot of $R_H$ as a function of $H_Z$ of a TbIG(25)|Pt(1.5)|Cu(1.5)|Tb$_{0.3}$Co$_{0.7}$(8) (Co-dominated) and spin valve. Plot of **g** $R_H$ and **h** $R$ as a function of $H_Z$ of a TbIG(25)|Cu(2)|Tb$_{0.55}$Co$_{0.45}$(8) (Tb-dominated) spin valve.

**SI4 – Temperature dependence**

The magnetic compensation temperature of TbIG can be determined by measuring the temperature dependence of the Anomalous Hall effect (AHE) in TbIG|Pt heterostructure, as shown in Ref.[12] and Fig. S4a. Here we observe a sign reversal of the AHE at approximately 200 K, which coincides with the maximum coercivity (Fig. S4b), due to the reduced Zeeman energy on the vanishingly small net magnetization. Similarly, from the magnetoresistance measurements of a TbIG|Cu|TbCo spin valve shown in Fig. S4c, we can determine the magnetic compensation of TbIG as the point at which the sign of the GMR reverses (see Fig. 3c of the main text) and the coercive field diverges. Above ~200K, the resistance ($R$) displays an upward jump when the magnetization of the TbIG layers reverses and a downwards jump in correspondence of the coercivity of TbCo. The amplitude of the second jump is larger than the first, as it is given by the sum of both the GMR effect and the MMR effect, which also leads to a decrease of $R$. Below the TbIG compensation ($TM$ ~200K) the GMR is negative, thus the resistance becomes lower when the magnetization of TbIG reverses. In this case, due to the GMR effect, we expect an increase of $R$ at the TbCo coercivity. However, here we observe a second jump of $R$ towards lower values. This is due to the sum of the GMR and MMR contributions, the latter being larger than the former, and thus resulting in a second decrease of $R$. Figure S4d summarizes the temperature dependence of the coercivity of the TbCo and TbIG layers inferred from the magnetoresistance measurement of Fig. S4c. While the $H_c$ of both layers increases gradually as the temperature is reduced, due to the stronger exchange interaction between the sublattices and enhanced PMA, the $H_C$ of TbIG shows an additional peak at $T_M = 200$ K. The $H_C$ vs $T$ trend for the TbIG layer obtained from AHE (Fig. S4b) and magnetoresistance (Fig. S4d) measurements is essentially the same.

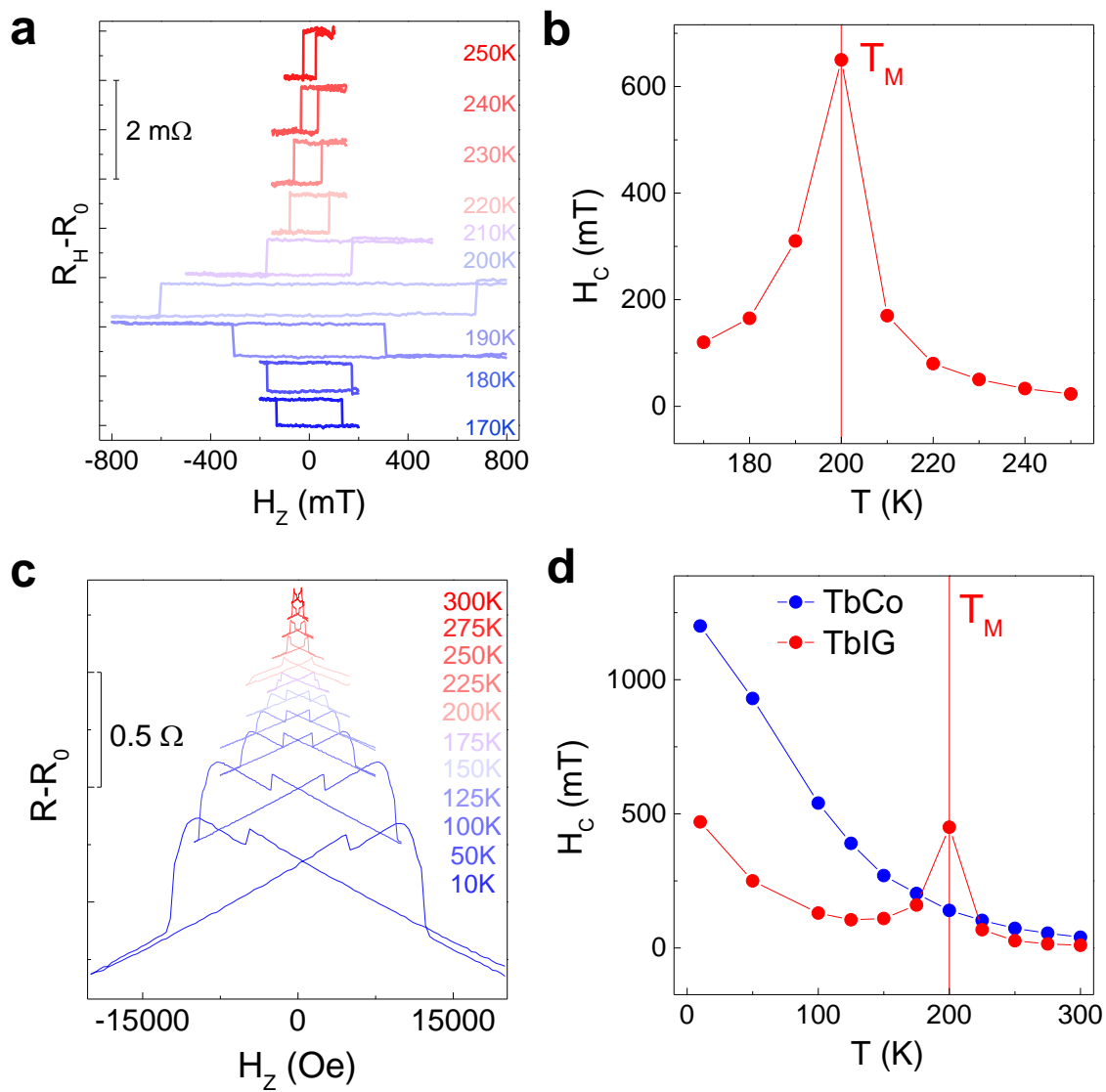

**Fig.S4 Temperature dependence. a** Plot of the Hall resistance ($R_H$) and as a function of out-of-plane magnetic field ($H_Z$) of a TbIG/Pt(4) bilayer measured between 170 and 250 K. **b** Plot of the coercive filed ($H_C$) of TbIG as a function of $T$ inferred from the AHE measurement of panel a. **c** Plot of the Resistance ($R$) and as a function of $H_Z$ of a TbIG(25)|Cu(2)|TbCo(8) spin valve measured between 10 and 300 K. **d** Plot of $H_C$ of TbIG (red) and TbCo (blue) as a function of $T$ inferred from the magnetoresistance measurement of panel c.

## SI5 – Theoretical model

We compute the conductivity of the trilayer by solving the following linearized Boltzmann transport equation with a layer-by-layer approach:

$$v_z \frac{\partial g}{\partial z} - \frac{eE}{m}\frac{\partial f_0}{\partial v_x} = -\frac{g}{\tau} \tag{S1}$$

Here $m$ and $\tau$ are the effective mass and momentum relaxation time of conduction electrons, $f_0$ is the equilibrium electron distribution function, $g(\mathbf{v}, z)$ is the deviation from $f_0$ induced by the external electric field with amplitude $E$ applied along **x**. Full translation symmetry is assumed in the x-y plane (i.e., the layer plane) so that the spatial dependence of $g$ only occurs in the thickness direction parallel to the z-axis. To further simplify our calculations, we assumed the same effective mass, relaxation time, and Fermi energy $\varepsilon_F$ for the conducting Cu and TbCo layers, which would be sufficient for the purpose of an order of magnitude estimation of the MR ratio of the trilayer system. The calculation can be extended straightforwardly to involve different values of the above-mentioned materials parameters without essential change to the formulation.

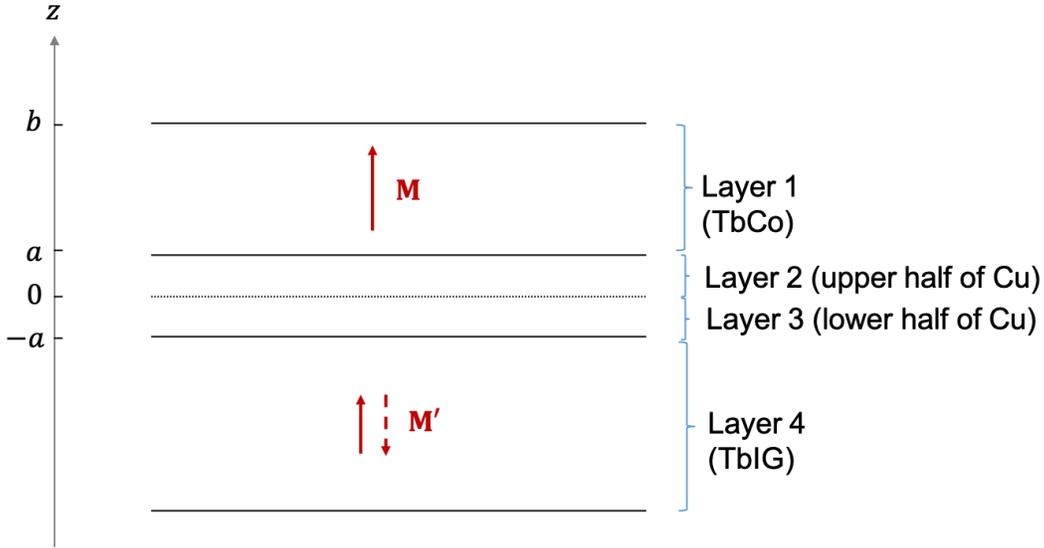

**Fig.S5 Coordinate system for calculating the magnetoconductivity in the TbCo|Cu|TbIG trilayer.** The TbCo and TbIG layers are denoted as Layer 1 and Layer 4, respectively, the Cu spacer is divided into two parts with equal thickness (a) by a dotted line where change of spin quantization axis occurs when the magnetizations of the two magnetic layers are antiparallel.

In order to capture the interfacial spin-dependent scattering, we divide $g$ into four additive components depending on the orientation of an electron's spin moment with respective to the magnetization of a magnetic layer and the sign of $v_z$ (i.e., the z-component of the electronic group velocity). The general solutions of Eq. (1) for each layer (as sketched in Fig. S5) can be written as

$$g^{(i)}_{\pm,\uparrow(\downarrow)}(\mathbf{v}, z) = \frac{eE\tau}{m}\frac{\partial f_0}{\partial v_x}\left[1 + C^{(i)}_{\pm,\uparrow(\downarrow)} exp\left(\frac{\mp z}{\tau|v_z|}\right)\right], \tag{S2}$$

where the "+" and "−" signs denote electrons with positive and negative $v_z$ components respectively, the arrow ↑ (↓) characterizes orientation of spin parallel (antiparallel) to the local magnetization of the magnetic layer in question, and the superscripts are the layer indices.

The twelve integration constants, $C^{(i)}_{\pm,\uparrow(\downarrow)}$ ($i = 1,2,3$) in Eqs. (S2) are determined by the following twelve boundary conditions at the interfaces between Layer 1, 2, and 3. For the ease of notation, we will suppress the group-velocity variable of $g^{(i)}_{\pm,\uparrow(\downarrow)}$.

- Reflection of electrons at the outer surface of the TbCo layer ($z = b$):

$$g^{(1)}_{-,\uparrow}(z = b) = R^{(b)}_{\uparrow} g^{(1)}_{+,\uparrow}(z = b) \tag{S3-1}$$

$$g^{(1)}_{-,\downarrow}(z = b) = R^{(b)}_{\downarrow} g^{(1)}_{+,\downarrow}(z = b) \tag{S3-2}$$

where $R^{(b)}_{\uparrow(\downarrow)}$ is the reflection parameter for spin-up (spin-down) electrons respectively. The value of $R^{(b)}_{\uparrow(\downarrow)}$ ranges from 0 to 1 with "0" corresponds to fully diffusive reflection and "1" corresponds to specular reflection.

- Spin-dependent electron reflection and transmission at the interface ($z = a$) between the TbCo and Cu layers:

$$g^{(1)}_{+,\uparrow}(z = a) = T^{(a)}_{\uparrow} g^{(2)}_{+,\uparrow}(z = a) + R^{(a)}_{\uparrow} g^{(1)}_{-,\uparrow}(z = a) \tag{S3-3}$$

$$g^{(1)}_{+,\downarrow}(z = a) = T^{(a)}_{\downarrow} g^{(2)}_{+,\downarrow}(z = a) + R^{(a)}_{\downarrow} g^{(1)}_{-,\downarrow}(z = a) \tag{S3-4}$$

$$g^{(2)}_{-,\uparrow}(z = a) = T^{(a)}_{\uparrow} g^{(1)}_{-,\uparrow}(z = a) + R^{(a)}_{\uparrow} g^{(2)}_{+,\uparrow}(z = a) \tag{S3-5}$$

$$g^{(2)}_{-,\downarrow}(z = a) = T^{(a)}_{\downarrow} g^{(1)}_{-,\downarrow}(z = a) + R^{(a)}_{\downarrow} g^{(2)}_{+,\downarrow}(z = a) \tag{S3-6}$$

where $R^{(a)}_{\uparrow(\downarrow)}$ and $T^{(a)}_{\uparrow(\downarrow)}$ are, respectively, the reflection and transmission parameters for spin-up (spin-down) electrons at the interface $z = a$.

- Change of spin quantization axis in the middle of the Cu layer ($z = 0$):

$$g^{(2)}_{+,\uparrow}(z = 0) = T_{\uparrow\uparrow}(\theta_m) g^{(3)}_{+,\uparrow}(z = 0) + T_{\uparrow\downarrow}(\theta_m) g^{(3)}_{+,\downarrow}(z = 0) \tag{S3-7}$$

$$g^{(2)}_{+,\downarrow}(z = 0) = T_{\downarrow\downarrow}(\theta_m) g^{(3)}_{+,\downarrow}(z = 0) + T_{\downarrow\uparrow}(\theta_m) g^{(3)}_{+,\uparrow}(z = 0) \tag{S3-8}$$

$$g^{(3)}_{-,\uparrow}(z = 0) = T_{\uparrow\uparrow}(\theta_m) g^{(2)}_{-,\uparrow}(z = 0) + T_{\uparrow\downarrow}(\theta_m) g^{(2)}_{-,\downarrow}(z = 0) \tag{S3-9}$$

$$g^{(3)}_{-,\downarrow}(z = 0) = T_{\downarrow\downarrow}(\theta_m) g^{(2)}_{-,\downarrow}(z = 0) + T_{\downarrow\uparrow}(\theta_m) g^{(2)}_{-,\uparrow}(z = 0) \tag{S3-10}$$

where $T_{\uparrow\uparrow}(\theta_m) = T_{\downarrow\downarrow}(\theta_m) = \cos(2\theta_m)$ and $T_{\uparrow\downarrow}(\theta_m) = T_{\downarrow\uparrow}(\theta_m) = \sin(2\theta_m)$ with $\theta_m$ the angle between the magnetizations of the TbCo and the TbIG layers.

- Reflection of electrons at the interface between the Cu and TbIG layers ($z = -a$):

$$g^{(3)}_{+,\uparrow}(z = -a) = R^{(-a)}_{\uparrow} g^{(3)}_{-,\uparrow}(z = -a) \tag{S3-11}$$

$$g^{(3)}_{+,\downarrow}(z = -a) = R^{(-a)}_{\downarrow} g^{(3)}_{-,\downarrow}(z = -a) \tag{S3-12}$$

Placing Eqs. (S2) into the above boundary conditions, one can determine the twelve unknowns, with which the spatially-averaged total conductivity of the trilayer structure can be calculated via

$$\sigma = \frac{-e}{(2\pi)^3} \sum_{i=1}^{3} \frac{1}{d_i} \int dz \int d^3 \mathbf{k} v_x \left( g_\uparrow^{(i)} + g_\downarrow^{(i)} \right) / E \tag{S4}$$

where $g_{\uparrow(\downarrow)}^{(i)} = g_{+,\uparrow(\downarrow)}^{(i)}$ for $v_z > 0$ and $g_{\uparrow(\downarrow)}^{(i)} = g_{-,\uparrow(\downarrow)}^{(i)}$ for $v_z < 0$. The longitudinal resistivity, $\rho$, can be obtained by inverting the conductivity tensor. And finally, the magnetoresistance ratio is obtained by

$$MR = \frac{\rho_{AP} - \rho_P}{\rho_P} \tag{S5}$$

where $\rho_{AP}$ and $\rho_P$ are the resistivities for the magnetizations of the two magnetic layers in the antiparallel ($\theta_m = \pi$) and parallel ($\theta_m = 0$) configurations, respectively. In Fig. S6, we plot the magnetoresistance ratio as a function of applied magnetic field for a trilayer TbIG(25 nm)|Cu(2 nm)|TbCo (8 nm), with the materials parameters given in Table S2. The calculated magnetoresistance ratio is 0.0062%, comparable to the value observed experimentally.

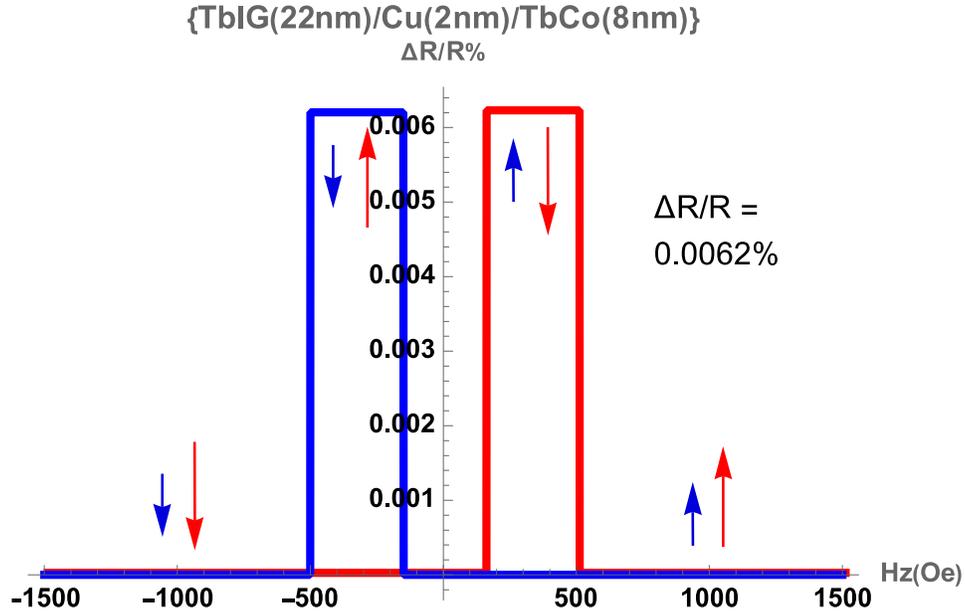

**Fig. S6 MR ratio as a function of the magnetic field applied perpendicular to the layer plane.** The blue (red) curve denotes the change of the MR with field sweep from +1500 Oe to -1500 Oe (from -1500 Oe to +1500 Oe). The red and blue arrows indicate, respectively, the magnetizations of the TbCo and TbIG layers, respectively. The coercive fields of TbIG and TbCo are taken to be 150 Oe and 500 Oe, respectively. And the materials used in the calculation are listed in the table below.

**Table S2 Reflection and Transmission parameters (Refs.[21],[45])**

| Symbols | Value | Parameter |
|---|---|---|
| $R_\uparrow^{(b)}$ | 0 | Reflection at x = b for spin-up electrons |
| $R_\downarrow^{(b)}$ | 0 | Reflection at x = b for spin-down electrons |
| $R_\uparrow^{(a)}$ | 0 | Reflection at x = a for spin-up electrons |
| $R_\downarrow^{(a)}$ | 0 | Reflection at x = a for spin-down electrons |
| $T_\uparrow^{(a)}$ | 0.5 | Transmission at x = a for spin-up electrons |
| $T_\downarrow^{(a)}$ | 0.95 | Transmission at x = a for spin-down electrons |
| $T_{\uparrow\uparrow}(\theta_m)$ | 0 for $\theta_m = \pi$, 1 for $\theta_m = 0$ | Spin-conserved transmission at x = 0 for spin-up electrons |
| $T_{\downarrow\downarrow}(\theta_m)$ | 0 for $\theta_m = \pi$, 1 for $\theta_m = 0$ | Spin-conserved transmission at x = 0 for spin-down electrons |
| $T_{\uparrow\downarrow}(\theta_m)$ | 1 for $\theta_m = \pi$, 0 for $\theta_m = 0$ | Spin-flip transmission (from spin-up to spin-down) at x = 0 |
| $T_{\downarrow\uparrow}(\theta_m)$ | 1 for $\theta_m = \pi$, 0 for $\theta_m = 0$ | Spin-flip transmission (from spin-down to spin-up) at x = 0 |
| $R_\uparrow^{(-a)}$ | 0.4995 | Reflection coefficient at x = −a for spin-up electrons |
| $R_\downarrow^{(-a)}$ | 0.5005 | Reflection coefficient at x = −a for spin-down electrons |